\documentstyle[epsfig,multicol,prl,aps]{revtex}
\begin{document}

\widetext

\title{
Tunneling Spectroscopy of the Underdoped High-$T_{c}$
Superconductors}

\author{Yunkyu Bang$^{1,2}$}
\address{
$^{1}$
Department of Physics, Chonnam National University,
Kwangju 500-757, Korea,}
\address{
$^{2}$
Center for Strongly Correlated Materials Research,\\
Seoul National University, Seoul 151-742, Korea.}
\author{Han-Yong Choi$^3$}
\address{
$^3$
Department of Physics and Institute for Basic Science Research,\\
Sung Kyun Kwan University, Suwon 440-746, Korea.}

\maketitle


\begin{abstract}
Considering a possibility that the pseudogap state in the
underdoped high-$T_{c}$ superconductors is not due to a
superconducting correlation, we study the tunneling spectroscopy
below $T_{c}$ based on the phenomenological models of
the pseudogap state.
Specifically we consider two cases: (1) the pseudogap is a
simple suppression of the density of states with an unknown
origin; (2) the pseudogap state is due to an antiferromagnetic
correlation. For both cases we calculate $dI/dV$ using the BTK
theory. The results are discussed in  comparison with
experiments.

\end{abstract}


\begin{multicols}{2}
\narrowtext

\section{Introduction}
The suppression of density of states around the Fermi level in the
underdoped high-$T_{c}$ superconductors (HTSC) above $T_{c}$ is
observed in various normal state experiments such as optical
conductivity, dc-resistivity, angle resolved photoemission, NMR,
tunneling spectroscopy, neutron scattering, specific heat,
Raman spectroscopy etc \cite{reviews},
and this phenomena is termed as
pseudogap (PG). Currently a consensus of the origin and
nature of this PG is still lacking and undoubtedly it is a key
issue to be resolved to make any progress toward a theory of high
Tc superconductivity.

About the origin of the PG state only two
possibilities are logically allowed.
The first possibility is that  the PG is somehow related with the
superconducting correlation and it develops into a real
superconducting gap below Tc.
And the second possibility is that the PG has nothing to do
with a superconducting gap but with something else. Along the
first line of thinking, preformed pair scenario \cite{Emery},
pairing fluctuations scenario \cite{Levin}, etc. are
proposed.
For the second possibility, antiferromagnetic
correlation\cite{Pines}, charge stripes\cite{Stripe}, etc
are considered as the origin of the PG.
At present experimental evidences exist both for superconducting
origin\cite{SC_origin} and for non-superconducting
origin\cite{NSC_origin,Deutscher,Krasnov}.
Therefore it is necessary to design an experiment to
identify  distinct features among the proposed scenarios.
Recently we proposed an experimental test using tunneling
spectroscopy in the PG state, specifically, for the preformed pair
scenario\cite{Choi}.
Namely we claim that there should be an Andreev
reflection signal even above Tc but below $T^*$ (PG cross-over
temperature) if there exist preformed Cooper pairs without long
range phase coherence. Until now there are only one positive \cite{LeeHJ} and
one negative \cite{Deutscher2}
experiment reported on the observation of an Andreev signal
in the PG region of the underdoped HTSC compounds.

In this paper we examine the second possibility for the PG state,
i.e., that the PG is irrelevant from the SC gap.
We calculate the tunneling
conductance ($dI/dV$) at zero temperature when the PG coexists with a SC
gap. Specifically we  considered two cases: (1) the PG is a simple
suppression of density of states of an unknown origin and the SC
correlation
has no direct interplay  with the PG;
(2) the PG is caused by an
AFM correlation (approximated by SDW) and below Tc the AFM
correlation and the SC correlation coexist and interplay with each
other. For both cases, we use the BTK theory\cite{BTK} to calculate the
tunneling conductance  ($dI/dV$). The results show characteristic
features of the tunneling density of states in each case and those
features can be used to sort out the true origin of the PG in
comparison with experiments.

The main results are: (1) for the first case, the basic line shape of the
tunneling conductance is a simple superposition of a standard BTK conductance and
an assumed PG density of states. When the SC gap size is smaller than
the PG size, the SC gap feature shows up as a
distinguishable peak inside the PG. However if the size of SC gap is
larger than the PG, the PG feature is overwhelmed by the SC gap
feature; (2) for the second case, there is an interesting
interplay  between the AFM and SC correlations.
Irrespective of the relative sizes
of the PG and SC gap, the tunneling conductance  shows only one
gap feature at $E=\Delta_{total}$ ($=\sqrt{\Delta_{SC}^2+\Delta_{PG}^2}$).
However
depending on the relative sizes of the PG and SC gap, the line shape
of $dI/dV$ looks very different. When the SC gap is bigger
than the PG, it looks more like a  conventional NIS
(Normal metal-Insulator-Superconductor) junction.
But for the other case
the main feature of the $dI/dV$ curve is determined by the SDW
correlation and shows  no diverging density of states approaching
the gap energy in contrast to the NIS junction. This
difference comes from the fact that the tunneling density of
states of the NISDW (Normal metal-Insulator-SDW state) junction
obtained by the BTK theory (even at large
barrier potential ($Z$) limit) is not the same as the actual
density of states, which would have been obtained by the tunneling
Hamiltonian method\cite{SDW_H}.
The difference of the tunneling conductance by
the BTK theory and by the tunneling Hamiltonian method in the SDW state,
which is in contrast to the NIS (Normal
metal-Insulator-Superconducor) junction, shows that the description
of tunneling  process by these two methods are not the same.
The important and interesting questions are then which description is
more proper description for actual tunneling process and  why in
the case of NIS  junction those two methods seem to give
consistent results for a $Z>>1$ limit.
More details will be discussed in
the later section.

\section{PG model I: simple suppression of density of states}

In this section we consider the case that the PG is a simple
suppression of density of states, which has been developed above $T_c$
by an unknown origin. An assumption is that the SC gap
develops below $T_c$ on top of the already developed PG density of states
and there
is no other correlation between the SC gap and the PG.

The tunneling conductance of NIS  junction is calculated by the BTK
theory as follows.

\begin{eqnarray}
dI(eV)/dV & = & 2 e v_F \it{A} \int^{\infty}_{-\infty} dE~ N_{PG}(E)
\frac{\partial f(E-V)}{\partial V} \nonumber \\
& & \cdot [1+|A(E)|^2-|R(E)|^2],
\end{eqnarray}

\noindent
 where
\begin{equation}
|A|^2 = \left\{ \begin{array}{ll} \frac{\Delta_S^2}{(E+(1+2 Z^2)
\sqrt{E^2-\Delta_S^2})^2}   & \mbox{for $E > \Delta_S$} \\

\frac{\Delta_S^2}{E^2+(1+2 Z^2)^2 (\Delta_S^2-E^2)} & \mbox{for $E
< \Delta_S$}
\end{array}
\right.
\end{equation}

\begin{equation}
|R|^2 = \left\{ \begin{array}{ll} \frac{4 Z^2
(E^2-\Delta_S^2)(Z^2+1)}{(E+(1+2 Z^2) \sqrt{E^2-\Delta_S^2})^2} &
\mbox{for $E > \Delta_S$} \\

\frac{4 Z^2 (Z^2+1) (\Delta_S^2-E^2)}{E^2+(1+2 Z^2)^2
(\Delta_S^2-E^2)} & \mbox{for $E < \Delta_S$}
\end{array}
\right.
\end{equation}

\noindent
are the coefficients for the Andreev and normal
reflections, respectively, and $Z$ is the strength of the
insulating barrier ($Z=\frac{mV}{\hbar^2 k_F}$).
The above equation
is a standard formula for the tunneling conductance by the BTK
theory with one modification, i.e., multiplied by the PG density of
states $N_{PG}(E)$. We simulate $N_{PG}(E)$ by Dynes' formula
with the SC gap replaced by the PG ($\Delta_P$).

\begin{equation}
N_{PG}(E) = 2 \pi N(0) Re \Big[ \frac{\omega +i
\Gamma}{\sqrt{(\omega+i \Gamma)^2-\Delta^2_P}} \Big].
\end{equation}

The numerical calculation has done for a tunneling into (1,0,0)
direction  toward the HTSC and assume that $\Delta_P$
and $\Delta_S$ has the maximum value at the same direction
(1,0,0). To simulate the surface roughness we also present the
angle averaged results by simply replacing $\Delta_{S,P}
\Rightarrow \Delta^0_{S,P} \cos (2 \theta)$. All the presented
tunneling conductance is normalized by the normal-state resistance
$R_N= (1+Z^2)/(2 N(0) e^2 v_F \it{A})$ and the bias voltage is
measured in unit of $\Delta_{P}$. Also for all the presented result we
chose $\Gamma=0.3$, which determines the shape of the PG density
of states $N_{PG}(E)$ from Eq(4).

\begin{figure}
\epsfig{figure=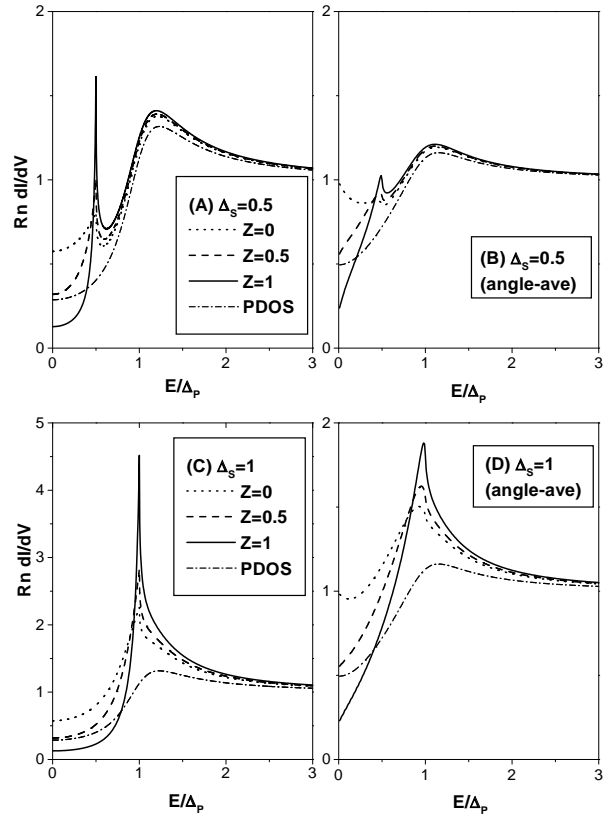,width=1.0\linewidth}
\caption{(a)
The normalized tunneling conductance $R_N dI/dV$
as a function of the bias voltage
in units of $\Delta_{P}$ with
$\Delta_{S}=0.5 \Delta_{P}$ for different Z (= 0, 0.5, 1).
$N_{PG}(E)$(PDOS) is also plotted for reference;
(b) The same as (a) but angle-averaged; (c) The same as (a) but
with $\Delta_{S}= \Delta_{P}$; (d) The same as (c) but
angle-averaged.
\label{fig1}}
\end{figure}

In Fig.1(a) we show the normalized tunneling conductance $R_N
dI/dV$ at zero temperature with the superconducting gap
$\Delta_{S}=0.5 \Delta_{P}$ for different barrier potentials
(Z=0,0.5, and 1). We also plot $N_{PG}(E)$ with
$\Gamma=0.3$ for comparison. The results can be trivially
understood. As well known from the BTK theory, for small value of
Z, we see the enhanced conductance below $\Delta_{S}$ due to the
Andreev reflection and this effect quickly disappears with
increasing Z value. The main difference of our results from the
conventional  BTK calculation is that we modulate the
conventional BTK tunneling conductance by multiplying with
$N_{PG}(E)$. As a result the enhanced conductance below $\Delta_{S}$ and the
sharp peak structure at $E=\Delta_{S}$ appear inside the PG
density of states ($\Delta_{S} < \Delta_{P}$).
In Fig.1(b) we show the same calculations but angle averaged to
simulated the surface roughness in real tunneling experiment.
The main features are the same as Fig.1(a) and the line shapes
become more rounded off. In particular the result of Z=1 case in
Fig.1(b) looks quite similar with the recent tunneling experiment
by Krasnov et al. \cite{Krasnov}

In Fig.1(c) and Fig.1(d) we show the results of same calculations
as Fig.1(a) and Fig.1(b), respectively  but with $\Delta_{S}=\Delta_P$.
When $\Delta_{S}$ has the same value as $\Delta_{P}$
the line shape of the tunneling conductance looks similar to
the conventional NIS junction, in particular with a large Z value.
The effect of
the PG is just to enhance the over-all line shape of the conductance.
For smaller Z value the conductance is still enhanced below $\Delta_{S}$
due to the Andreev reflection.
However even the enhanced
conductance  at $E \rightarrow 0$ limit is far below than 2
(for a conventional
NIS junction it approaches 2 for  $E < \Delta_{S}$
as Z $\rightarrow$ 0) because of the reduced DOS of PG origin.
Again the angle
averaged results in Fig.1(d) shows a more rounded off line
shapes. The angle averaged PG density of states (PDOS)
is shown for comparison in Fig.1(b) and (d).

 In Fig2.(a-d), we plot the normalized conductance with varying
 $\Delta_{S}(=$0.5,1, and 1.5 $\Delta_{P}$).
 When Z=1 (Fig.2(a) and Fig.2(b)) the results are easily
 understood. For $\Delta_{S} < \Delta_{P}$ the distinct peak
 structure due to SC appears inside the PG as explained above.
 And when $\Delta_{S} > \Delta_{P}$ there is no more distinct peak
 structure and the SC feature overwhelms the PG structure.
 When Z=0.1 (Fig.2(c) and Fig.2(d)) the conductance line shapes look
 more peculiar, but basically it can be understood as an overlap of the
 Andreev enhanced conductance below $\Delta_{S}$ and the PG
 density of states $N_{PG}(E)$. G. Deutscher et al.
 \cite{Deutscher} reported that there are two energy scales observed
 in tunneling experiments, and depending on the barrier strength
 ($Z$) only one of the two energy scales dominates the conductance line shape.
 In Fig2.(c) the line with $\Delta_{S}=0.5\Delta_{P}$ (solid line) shows a
 similar feature to the experimental data in Fig.2 of Ref.\cite{Deutscher}.

 In view of the  experiments of Krasnov et al.
 \cite{Krasnov} and G. Deutscher et al.
 \cite{Deutscher}, if our
 calculations have any relevance with underdoped HTSC, it would be
 the case of $\Delta_{S} < \Delta_{P}$ (Fig.1(a)(b) and Fig.2(c)(d))
 and it
 will be interesting if more tunneling experiments with different Z
 parameters become available in near future.

 \section{PG model II:SDW}

 In this section we consider another possibility of the PG,
 namely, in which  the correlation, which induces the PG, coexists
 with a SC correlation below
 $T_c$ and the PG correlation interplay with the SC correlation.
 In this case the key difference from the case (I) is that either
 $\Delta_{S}$ or $\Delta_{P}$ does not show up as   separate
 features  in  the
 tunneling conductance but the total gap $\Delta_{total}=\sqrt
 {\Delta_{S}^2 + \Delta_{P}^2}$ shows up as a result of the
 interplay of two gaps. However more detailed line shape of the
 tunneling conductance reveals the existence of  two gaps,
  $\Delta_{S}$ and $\Delta_{P}$.

\begin{figure}
\epsfig{figure=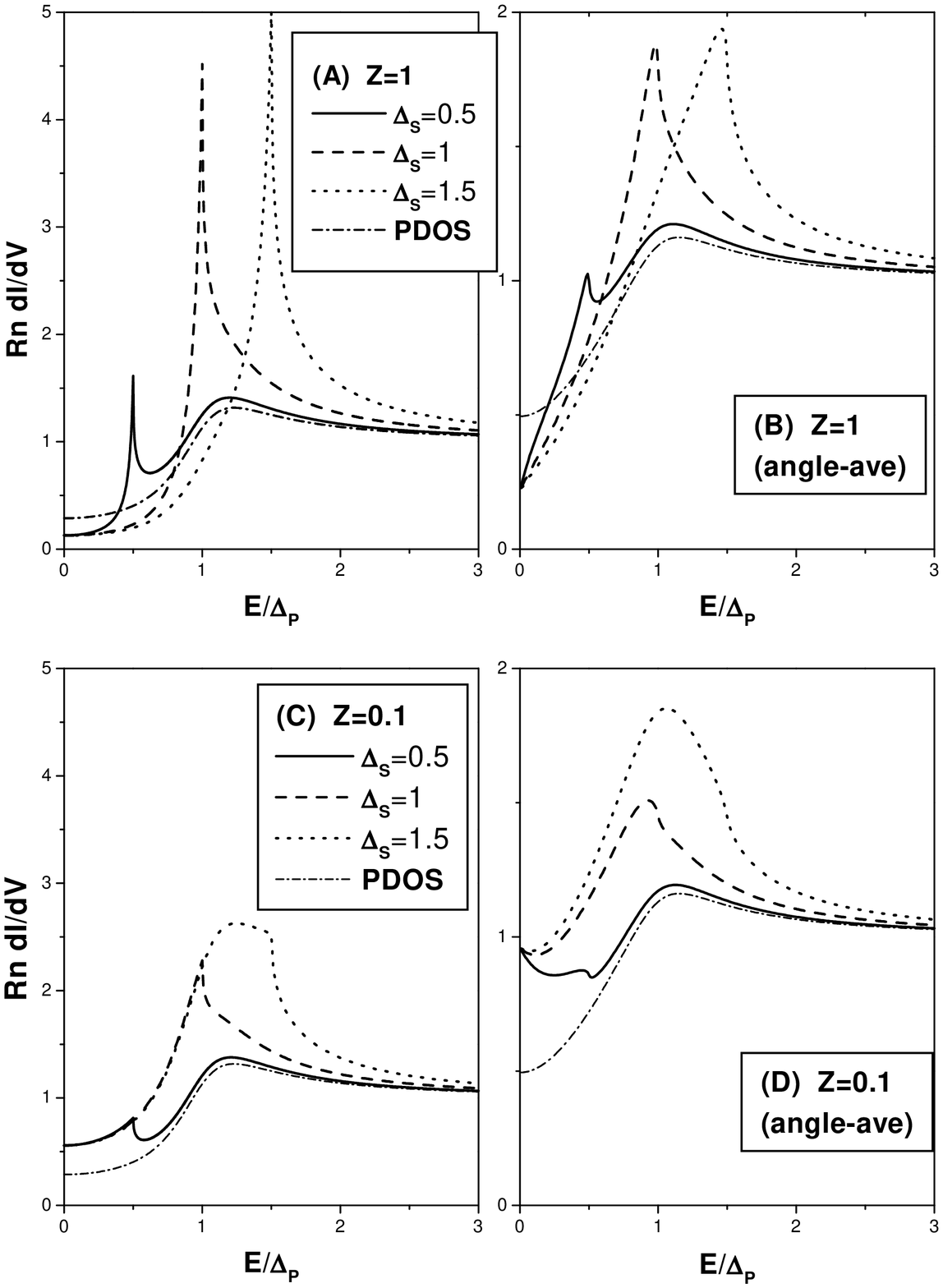,width=1.0\linewidth}
\caption{(a)
The normalized tunneling conductance $R_N dI/dV$
as a function of the bias voltage
in units of $\Delta_{P}$ with  Z=1for different $\Delta_{S}$ (= 0.5, 1, 1.5 $\Delta_{P}$).
$N_{PG}(E)$(PDOS) is also plotted for reference;
(b) The same as (a) but angle-averaged; (c) The same as (a) but
with Z=0.1; (d) The same as (c) but
angle-averaged.
\label{fig2}}
\end{figure}

 Specifically, we assume the SDW correlation as the
 origin of PG. The reason for this assumption is of course that there
 is a strong antiferromagnetic (AFM) correlation in underdoped
 HTSC compounds. Particularly for the underdoped HTSC compounds
 there is  clear experimental evidence
 from neutron scattering \cite{neutron} for the coexistence of the
 AFM correlation and superconductivity  below $T_c$ and also the
 AFM correlation length becomes much larger than the SC
 correlation length. This enable us to treat the short range AFM
 correlation by SDW from the point of view of the SC correlation.

\subsection{tunneling with SDW only}

As a prelude we
first generalize the BTK theory to the NISDW
 (Normal metal-Insulator-SDW) junction. For simplicity we consider
 only a one dimension model and assume a commensurate SDW state
 with $Q=k_{F}=\pi/a$ ($a$ is the lattice distance) and  also neglect the
 Fermi surface mismatch between the normal metal and the SDW state.

 The wave functions of the lefthand (normal metal) and the righthand (SDW)
 sides of the tunneling barrier is written as,

\begin{eqnarray}
\Psi_L (x) & = & e^{i kx} + R e^{- i kx}  \nonumber \\
\Psi_R (x) & = & T [a(k) e^{-i kx} + b(k) e^{- i (k-Q)x}]
\end{eqnarray}

\noindent
where $a(k)^2=\frac{\epsilon+\xi}{2 \epsilon}$ and
$b(k)^2=\frac{\epsilon-\xi}{2 \epsilon}$ are the Bogoliubov coefficients of
the SDW state,  $\epsilon= \sqrt{\xi^2 + \Delta_{SDW}^2}$, and
$\xi=\hbar^2 k^2/2m -\mu$.
By matching the boundary conditions of the wave functions at
$x=0$, we obtained $R$ and $T$ as follows.

\begin{eqnarray}
R(E) & = & -\frac{(a+b) i Z -a}{(a+b) i Z +b} \noindent \\
T(E) & = & \frac{1}{(a+b) i Z + b}
\end{eqnarray}

For $\epsilon^2 > \Delta_{SDW}^2$, $|R|^2$ is easily calculated,
and for $\epsilon^2 < \Delta_{SDW}^2$, using $a^2=\frac{\epsilon-i
\sqrt{\Delta_{SDW}^2-\epsilon^2}}{2 \epsilon}$ and $b^2=\frac{\epsilon + i
\sqrt{\Delta_{SDW}^2-\epsilon^2}}{2 \epsilon}$, we can show
$|R|^2=1$.

The tunneling conductance of NISDW junction with different
Z (=0,0.3, and 1) are shown in Fig.3(a,b,c) as dash-dot lines
($\Delta_{S}=0$ case). The tunneling conductance line shape is
qualitatively different from the actual density of states
$N_{SDW}(E)$, which would have been obtained by the tunneling
Hamiltonian method.
Here we clearly observe that the BTK theory and the
tunneling Hamiltonian method give different results for tunneling
conductance. And it has been already known that these two methods describe
different physical processes for the tunneling phenomena
\cite{BTK}. In particular, the tunneling Hamiltonian method,
although  physically appealing, has never been justified in
regard with the tunneling transfer matrix element $T$ (it is always
taken as a constant).
Then another question may arise: why then for the BCS state
does the BTK theory give a qualitatively similar result
as the tunneling Hamiltonian methods
at least for a large $Z$ limit where the Andreev reflection
process is suppressed, but not for the SDW state ?
The reason is that for the BCS state the quasiparticle is a
superposition of a momentum $k$ particle and a momentum $-k$ hole,
and all together it carries the flux of the same momentum $k$, while
for the SDW state the quasiparticle is a superposition  of momenta $k$ and
$k+Q$ particles and therefore the quasiparticle in SDW state
doesn't carry a single momentum.
For the BTK theory of tunneling
the main physics is the flux conservation described by Liouville's
theorem, therefore the BTK theory sensitively traces the correlation of
different momentum states in the SC or SDW states, while the tunneling
Hamiltonian method is completely blind of the momentum correlation.
In the case of
SC (BCS) state, we are in  a fortune situation as described above,
therefore  momentum correlation doesn't play an important role
but particle-hole correlation plays an important role in the BTK
theory, which provides the main difference between the BTK theory
and the tunneling Hamiltonian method through the Andreev scattering.
Now for the case of SDW state, the momentum correlation is
important  but not the particle-hole correlation;
therefore it is expected that the two methods will give different
results.

An important question is then which theory should be trusted.
As described above, the tunneling Hamiltonian method has never been
justified, while there is no logical flow in the BTK theory based
on the flux conservation. Therefore we think unless the tunneling
interface is very rough (of course we need to estimate how much
rough is rough) the BTK theory becomes more trustable.

\subsection{tunneling with SDW+SC}

Now we will consider the tunneling junction of a normal
metal-insulator-SDW+SC (NISDWS). Again for simplicity we consider
only a one dimensional model. In reality, as in HTSC, the analysis of
tunneling in the two dimensional SDW state is more complicated, in
particular when SDW+SC is considered.
However our one dimensional model is enough to provide a qualitative
understanding of  the
interplay of SDW and SC correlations in tunneling process.

Once the SDW state is formed, there are two branches of
quasiparticles created: $\alpha_{+,k}=a(k) c_{k}+ b(k) c_{k+Q}$ and
$\alpha_{-,k}=b(k) c_{k}- a(k) c_{k+Q}$, where $a(k)$ and $b(k)$
are the Bogoliubov coefficients as defined above.
Now we assume that the superconducting pairing occurs between
$\alpha_{+,k}$ and $\alpha_{+,-k}$, and also between
$\alpha_{-,k}$ and $\alpha_{-,-k}$, but not between
$\alpha_{+,k}$ and $\alpha_{-,-k}$, for example.
Although this assumption is mainly for simplicity of the analysis,
it is known that the  pairing interaction between different branches are much
weaker in Hubbard model\cite{SWZ}.
Also strictly speaking the SDW state with a commensurate wave vector
($Q=\pi/a$) is an insulator with a fully developed gap below
$\Delta_{SDW}=\Delta_{P}$,
in contrast to the PG state in
HTSC where  residual density of states still remain and the system
remains a metal.
Therefore it is clear that our SDW state only mimic a state
with a short range AFM correlation of the underdoped HTSC and it is
justified by the fact that  the AFM correlation
is effectively  long range from the viewpoint  of the SC
correlation.
A main drawback of this assumption is that the effect
of the residual density of states below PG in the tunneling
conductance is completely missing.
In summary, the purpose of this section is to study a tunneling
characteristics of a SDW+SC state due to the interplay of two
correlations, but not all the details of  real materials.

Now for the BTK theory the wave functions of the lefthand (normal metal)
and righthand (SDW+SC) sides of the tunneling interface are
written as,

\begin{eqnarray}
\Psi_{L}(E) & = & \left( \begin{array}{c}
1 \\ 0 \end{array} \right) e^{i kx} +
R(E) \left( \begin{array}{c}
1 \\ 0 \end{array} \right) e^{-i kx} +
A(E) \left( \begin{array}{c}
0 \\ 1 \end{array} \right) e^{i kx}, \nonumber \\
\Psi_{R}(E) & = &  C(E) \left( \begin{array}{c}
u \\ v \end{array} \right) \alpha_{+}(-k) +
D(E) \left( \begin{array}{c}
v \\ u \end{array} \right) \alpha_{-}(-k).
\end{eqnarray}

\noindent
where $E^2=\sqrt{\epsilon^2 + \Delta_{S}^2}=
\sqrt{\xi^2 + \Delta_{SDW}^2 + \Delta_{S}^2}$ and
$\xi=\hbar^2 k^2/2m -\mu$. All $k \approx k_{F}$ approximation
is taken; the error of the approximation is
$O(\Delta_{total}/E_{F})$.
\noindent
Also a shorthand notation of the wave functions are

\begin{eqnarray}
\alpha_{+}(-k) & = & a(\epsilon) e^{-i kx} + b(\epsilon) e^{-i
(k-Q)x}, \nonumber \\
\alpha_{-}(-k) & = & b(\epsilon) e^{-i kx} - a(\epsilon) e^{-i
(k-Q)x}.
\end{eqnarray}

\noindent
As usual the SC Bogoliubov coefficients are
$u(E)^2 = (E+\epsilon)/2E$ and $u(E)^2 = (E-\epsilon)/2E$.
By matching boundary conditions, which is a lengthy but straight
forward calculation, we obtained the coefficients of $A(E)$ and $R(E)$ as
follows.

\begin{eqnarray}
A & =& \frac{u v}{F},\nonumber \\
R & =& \frac{(u^2-v^2)[Z^2 (b^2-a^2)+ i Z (b^2-a^2+2ab) -a b]}{F},
\nonumber \\
F &= & (u^2-v^2)[Z^2 (b^2-a^2)+ i Z \cdot 2 a b]+b^2 u^2 + a^2 v^2.
\end{eqnarray}

\noindent
In order to calculate $|A|^2$ and $|R|^2$ we need to divide the
region of $E$ into three regions.
For region (I), where $E > \Delta_{total} (=\sqrt{\Delta_{P}^2+ \Delta_{S}^2}$),
$\epsilon=\sqrt{E^2-\Delta_{S}^2}$ and
$\xi=\sqrt{E^2-\Delta_{total}^2}$.
For region (II), where $\Delta_{S} <E < \Delta_{total}$,
$\epsilon=\sqrt{E^2-\Delta_{S}^2}$ and
$\xi= i \sqrt{\Delta_{total}^2 - E^2}$, and finally
for region (III), where $E < \Delta_{S}$,
$\epsilon= i \sqrt{\Delta_{S}^2 - E^2}$ and
$\xi= i \sqrt{\Delta_{total}^2 - E^2}$.
But in the final results of $|A|$ and $|R|$, $\epsilon$ factors
are all cancelled out, so that the region (II) and (III) are not
distinguished. Therefore the tunneling characteristics shows
distinct change only across $E=\Delta_{total}$ but no change
across either $E=\Delta_{S}$ or $E=\Delta_{P}$.

The final results of the Andreev ($A$) and normal reflection ($R$)
coefficients are: for region (I),

\begin{eqnarray}
|A|^2 & =& \frac{\Delta_{S}^2}{G}, \nonumber \\
|R|^2 & =& \frac{(2 Z^2
\sqrt{E^2-\Delta_{t}^2}-\Delta_{P})^2+4Z^2(\sqrt{E^2-\Delta_{t}^2}+\Delta_{P})^2}{G},
\nonumber \\
G &=& [(2Z^2+1)\sqrt{E^2-\Delta_{t}^2}+E]^2 + 4Z^2 \Delta_{P}^2;
\end{eqnarray}

\noindent
and  for region (II) and (III),

\begin{eqnarray}
|A|^2 & =& \frac{\Delta_{S}^2}{H}, \nonumber \\
|R|^2 & =& \frac{(\Delta_{P}+2Z \sqrt{\Delta_{t}^2-E^2})^2 + 4 Z^2
(Z \sqrt{\Delta_{t}^2-E^2} + \Delta_{P})^2}{H}, \nonumber \\
H &=& E^2 + [(2Z^2+1) \sqrt{\Delta_{t}^2-E^2}+ 2 Z \Delta_{P}]^2.
\end{eqnarray}

\begin{figure}
\epsfig{figure=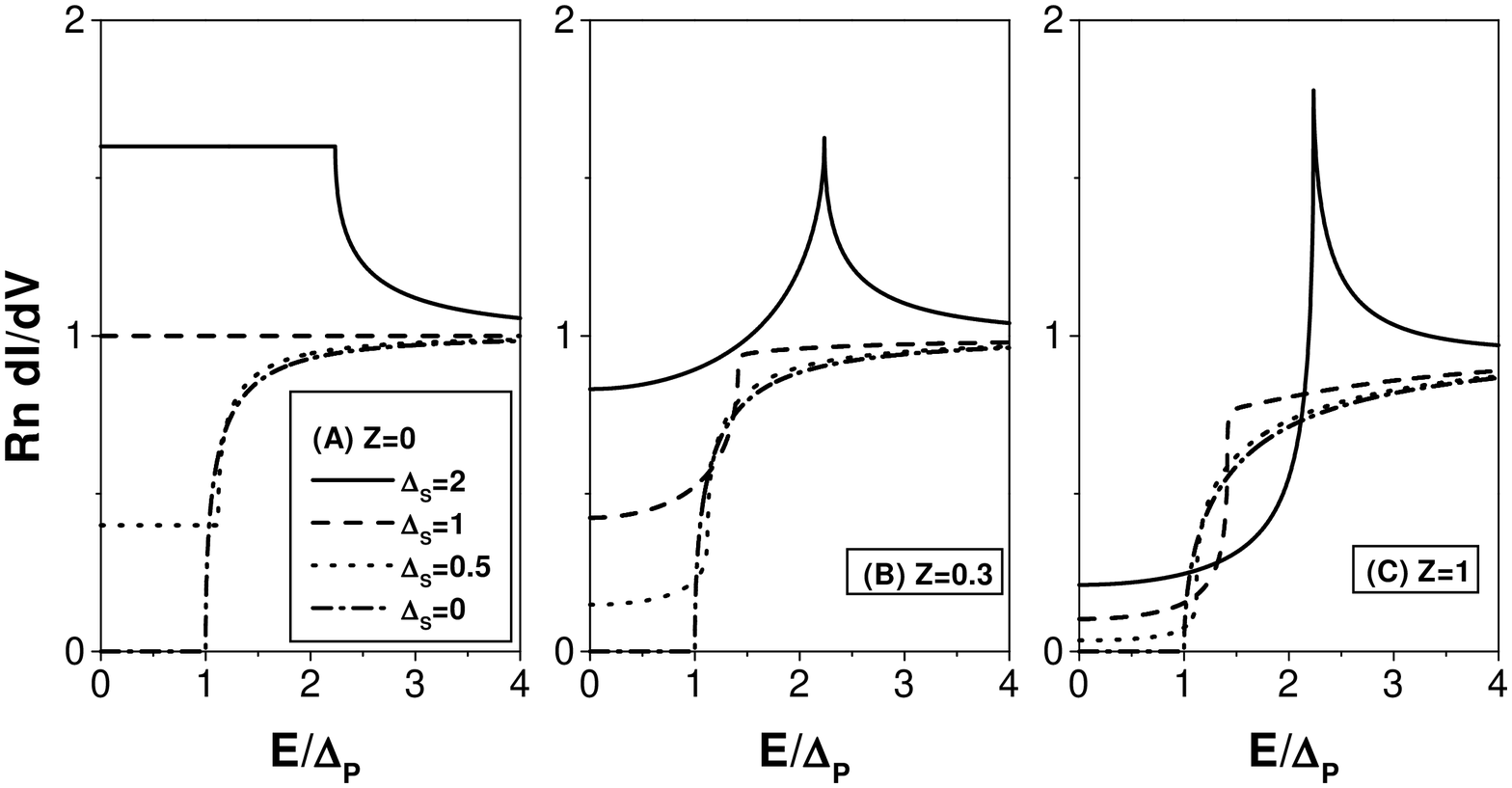,width=1.1 \linewidth}
\caption{(a)
The normalized tunneling conductance $R_N dI/dV$ for NISDWS
(Normal metal-Insulator-SDW+SC) junction
as a function of the bias voltage
in units of $\Delta_{P}$  with Z=0
for different $\Delta_{S}$ (=0, 0.5, 1, 2 $\Delta_{P}$);
(b) The same as (a) but with Z=0.3; (c) The same as (a) but
with Z=1.
\label{fig3}}
\end{figure}

In Fig.3(a-c) we plot the numerical calculations of the
normalized  tunneling conductance as a function of bias voltage
(in unit of $\Delta_{P}$) with varying size of $\Delta_{S}$
(=0,0.5,1, and 2) for different barrier potentials (Z=0, 0.3, and 1).
The main features of the tunneling conductance are very different from
a pure  SC case.
When $\Delta_{S}$ is much bigger than $\Delta_{P}$
(say, $\Delta_{S}/\Delta_{P}=2$ in Fig.3) the line shape is
similar to the pure  SC case.
But even in this case the position of a gap in the conductance
is determined by $\Delta_{t}$ and the presence of SC gap
$\Delta_{S}$ only show up through an Andreev scattering
coefficient, which enhances the conductance below the total gap.
On the other hand when $\Delta_{S} \leq \Delta_{P}$,
the line shape looks closer to the pure
SDW case, which is qualitatively different from a real density of
states as explained in the previous section. Again  the presence
of SC gap $\Delta_{S}$ show up only through an Andreev scattering
and  enhances the conductance below the total gap.

\begin{figure}
\epsfig{figure=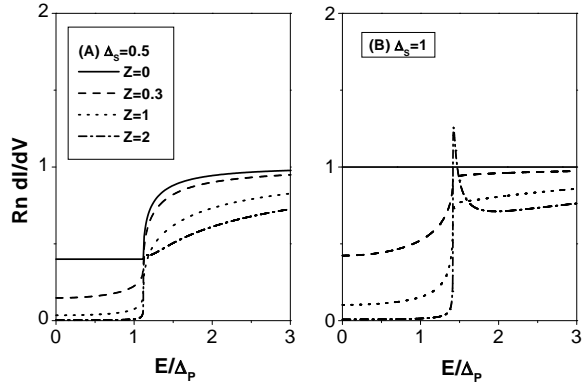,width=1.0 \linewidth}
\caption{(a)
The normalized tunneling conductance $R_N dI/dV$ for NISDWS
(Normal metal-Insulator-SDW+SC) junction
as a function of the bias voltage
in units of $\Delta_{P}$ with $\Delta_{S}=0.5 \Delta_{P}$
for different Z (=0, 0.3, 1, 2);
(b) The same as (a) but with $\Delta_{S}= \Delta_{P}$
.\label{fig4}}
\end{figure}

In Fig.4(a)(b) we plot the same calculations in different
organization. Here we vary Z values with fixed $\Delta_{S}$ to see
more clearly
the effect of varying strength of the  tunneling barrier. Main
features are already explained above. Both in Fig.4(a) and (b),
the normalized conductance below $\Delta_{total}$ is enhanced by
the Andreev reflection and it increases as $Z \rightarrow 0$. When
Z=0, we can have a simple results of $|A|^2=\frac{\Delta_{S}^2}{\Delta_{t}^2}$ and
$|R|^2=\frac{\Delta_{P}^2}{\Delta_{total}^2}$ for $E < \Delta_{total}$.
These expressions of $|A|^2$ and $|R|^2$ clearly show the
contrasting role of the SDW gap (normal reflection) and the SC gap
(Andreev reflection); the normalized conductance is given by $R_N
dI/dV (E < \Delta_{t})=\frac{2 \Delta_{S}^2}{\Delta_{t}^2}$ in this limit.
Increasing Z from zero, it continues to suppress the  Andreev
reflection and enhances the normal reflection.

In view of current tunneling experiments, if our results of the
NISDWS junction has any relevance with the underdoped HTSC, only
possibility is the case of $\Delta_{S} > \Delta_{P}$ at low
temperature. In order to realize this possibility starting with
a sizable $\Delta_{P}$ and  a zero $\Delta_{S}$ above $T_{c}$, we
have to imagine that $\Delta_{P}$ gradually decreases and $\Delta_{S}$
gradually increases as temperature is lowered.
Because this scenario is quite unlikely from the currently available
experiments, we can conclude either that the PG is not caused by
an AFM correlation, or that even if the origin of PG is an AFM correlation
the AFM correlation is not strong enough to have any significant
interplay with a SC correlation.

\section{Conclusion}

We considered two phenomenological models of the PG state in underdoped HTSC
and studied the characteristics of tunneling conductance at zero
temperature by generalizing the BTK theory.
In model I, we assumed that the PG is a simple suppression of
density of states of a unknown origin and when the system goes to
SC state there is no direct interplay between the PG correlation
and the SC correlation. In this case the characteristics of tunneling conductance
is a simple superposition of a standard tunneling conductance with
SC gap $\Delta_{S}$ and the PG density of states $N_{PG}(E)$.
Despite the simplicity of the model, the results seem to explain
some of the recent tunneling experiments by Krasnov et al.\cite{Krasnov}
and G. Deutscher \cite {Deutscher},
which indicate non-superconducting origin of the PG.
In the model II, we assumed that the PG is caused by an AFM
correlation and it is simulated by SDW state. Below $T_{c}$ the SDW
correlation and SC correlation show an interesting interplay. As a
result, the tunneling gap is given by
$\Delta_{total}=\sqrt{\Delta_{S}^2+\Delta_{P}^2}$ and individual
gaps, $\Delta_{S}$ and $\Delta_{P}$, do not show up explicitly.
In particular when $\Delta_{P,SDW} > \Delta_{S}$, the line shape
of the tunneling conductance looks qualitatively different from a
conventional NIS junction. In view of available tunneling experiments in
underdoped HTSC, the relevance of the model II is  possible
only when $\Delta_{P,SDW} < \Delta_{S}$ below $T_{c}$, which is quite
unlikely at present. To clarify the issue of the PG  more
experiments on tunneling  are essential
and our study should serve
as a useful benchmark.

We would like to thank H.J. Lee for invaluable
discussion of his data and G. Deutscher for sending us their
preprint prior to publication, respectively.
This work was supported
by the Korean Science and Engineering Foundation (KOSEF)
through the Center for Strongly Correlated Materials Research (CSCMR)
(2000)(YB) and through the Grant No. 1999-2-114-005-5 (YB and
HYC).

\end{multicols}


\begin{references}

\bibitem{reviews} For a review, see, for example,
T. Timusk and B. Statt, Rep. Prog. Phys. {\bf 62}, 61
(1999); M. Randeria, cond-mat/9710223.

\bibitem{Emery} V. J. Emery and S. A. Kivelson, Nature {\bf 374}, 434
(1995).

\bibitem{Levin}
 J. R. Engelbrecht, A. Nazarenko, M. Randeria, and
E. Dagotto, Phys. Rev. B {\bf 57}, 13406 (1998);
Q. Chen, I. Kosztin, B. Janko, and K. Levin, Phys.
Rev. Lett. {\bf 81}, 4708 (1998).

\bibitem{Pines} D. Pines, Physica C {\bf 282-287}, 273 (1997);
 A. V. Chubukov, D. Pines, and B. P. Stojkovic, J.
Phys. Condens. Matter {\bf 8}, 10017 (1996); A. V. Chubukov and J.
Schmalian, Phys. Rev. B {\bf 57}, R11085 (1998).

\bibitem{Stripe}
C. Castellani et al., cond-mat/0001231;
R.S. Gonnelli et al., cond-mat/0003100.


\bibitem{SC_origin}
Ch. Renner et al.,  Phys. Rev. Lett. {\bf 80}, 149 (1998);
N. Miyakawa, et al., Phys. Rev. Lett. {\bf 83}, 1018 (1999).

\bibitem{NSC_origin}
J.L.Tallon, cond-mat/9911422.

\bibitem{Deutscher}
G. Deutscher, Nature {\bf 397} 410 (1999).

\bibitem{Krasnov}
V.M. Krasnov, A.Yurgens, D. Winkler, P.Delsing
and T.Claeson, cond-mat/0002172.

\bibitem{Choi}
H.Y. Choi, Y. Bang, and D. K. Campbell, cond-mat/9902125 (to appear in
Phys. Rev. B).

\bibitem{LeeHJ}
H.J. Lee et al., (private communication).

\bibitem{Deutscher2}
G. Deutscher et al., preprint.

\bibitem{BTK}
G.E.Blonder, M.Tinkham, and T.M.Klapwijk, Phys. Rev. B {\bf 25}, 4525
(1982).

\bibitem{SDW_H}
A.M. Gabovich and A.I. Voitenko, Phys. Rev. B {\bf 52}, 7437
(1995).

\bibitem{neutron}
Ch. Niedermayer et al.,  Phys. Rev. Lett. {\bf 80}, 3843 (1998);
K. Yamada et al., Phys. Rev. B {\bf 57}, 6165 (1998);
Y.S. Lee et al., cond-mat/9902157.

\bibitem{tunneling_H}
J. Bardeen,  Phys. Rev. Lett. {\bf 6}, 57 (1961);
M.H. Cohen, L.M. Falicov, and J.C. Phillips,  Phys. Rev. Lett. {\bf 8}, 316
(1962).

\bibitem{SWZ}
J.R.Schrieffer,X.G. Wen and S.C. Zhang, Phys. Rev. B {\bf 39},
11663, (1989).







\end{references}
\end{document}